\documentclass{aastex62}

\usepackage{hyperref}
\usepackage{multirow}
\usepackage{amsmath}
\usepackage{graphicx} 

\shorttitle{Predicting Solar Flares Using a Long Short-Term Memory Network}
\shortauthors{Liu et al.}

\begin{document}

\title{{\bf \large Predicting Solar Flares Using a Long Short-Term Memory Network}}

\author{Hao Liu}
\affiliation{Institute for Space Weather Sciences, New Jersey Institute of Technology, University Heights, Newark, NJ 07102-1982, USA hl422@njit.edu, chang.liu@njit.edu, wangj@njit.edu, haimin.wang@njit.edu}
\affiliation{Department of Computer Science, New Jersey Institute of Technology, University Heights, Newark, NJ 07102-1982, USA}

\author{Chang Liu}
\affiliation{Institute for Space Weather Sciences, New Jersey Institute of Technology, University Heights, Newark, NJ 07102-1982, USA hl422@njit.edu, chang.liu@njit.edu, wangj@njit.edu, haimin.wang@njit.edu}
\affiliation{Big Bear Solar Observatory, New Jersey Institute of Technology, 40386 North Shore Lane, Big Bear City, CA 92314-9672, USA}
\affiliation{Center for Solar-Terrestrial Research, New Jersey Institute of Technology, University Heights, Newark, NJ 07102-1982, USA}

\author{Jason T. L. Wang}
\affiliation{Institute for Space Weather Sciences, New Jersey Institute of Technology, University Heights, Newark, NJ 07102-1982, USA hl422@njit.edu, chang.liu@njit.edu, wangj@njit.edu, haimin.wang@njit.edu}
\affiliation{Department of Computer Science, New Jersey Institute of Technology, University Heights, Newark, NJ 07102-1982, USA}

\author{Haimin Wang}
\affiliation{Institute for Space Weather Sciences, New Jersey Institute of Technology, University Heights, Newark, NJ 07102-1982, USA hl422@njit.edu, chang.liu@njit.edu, wangj@njit.edu, haimin.wang@njit.edu}
\affiliation{Big Bear Solar Observatory, New Jersey Institute of Technology, 40386 North Shore Lane, Big Bear City, CA 92314-9672, USA}
\affiliation{Center for Solar-Terrestrial Research, New Jersey Institute of Technology, University Heights, Newark, NJ 07102-1982, USA}

\begin{abstract}
We present a long short-term memory (LSTM) network for  
predicting whether an active region (AR) would produce a 
$\Upsilon$-class flare within the next 24 hours.
We consider three $\Upsilon$ classes, namely
$\ge$M5.0 class, $\ge$M class, and $\ge$C class, and
build three LSTM models separately, 
each corresponding to a $\Upsilon$ class.
Each LSTM model is used to make predictions of its corresponding $\Upsilon$-class flares.
The essence of our approach is to model data samples in an AR as time series and
use LSTMs to capture temporal information of the data samples. 
Each data sample has 40 features including 25 magnetic parameters 
obtained from the Space-weather HMI Active Region Patches (SHARP) 
and related data products as well as 15 flare history parameters.
We survey the flare events that occurred from 2010 May to 2018 May, 
using the \textit {GOES} X-ray flare 
catalogs provided by the 
National Centers for Environmental 
Information (NCEI), 
and select flares 
with identified ARs in the NCEI flare catalogs.
These flare events are used to build the labels (positive vs. negative) of the data samples.
Experimental results show that (i) using only 14-22 most important features 
including both flare history and magnetic parameters
can achieve better performance than using all the 40 features together;
(ii) our LSTM network outperforms related machine learning methods
in predicting the labels of the data samples.
To our knowledge, this is the first time that LSTMs have been used for solar flare prediction.
\end{abstract}


\section{Introduction} \label{sec:intro}

Solar flares, the largest explosive events in our solar system,    
are intense bursts of radiation that occur 
in the Sun's atmosphere
and release massive amounts of energy into space.
They last from minutes to hours and are often seen as 
bright chromospheric ribbons and hot coronal loops on the Sun. 
Some flares are small and innocent while others can be tremendous and violent. 
Powerful flares and the often accompanied 
coronal mass ejections (CMEs) 
can cause severe influences on the near-Earth environment, 
resulting in potentially 
life-threatening consequences \citep{Daglis04}. 
Therefore, substantial efforts are being invested
on solar flare research including forecast and mitigation plans. 

The triggering mechanism of solar flares is far from being fully understood.
Many studies have shown that flares and CMEs could be powered 
by the free magnetic energy accumulated in the coronal field, 
which can be impulsively released by magnetic reconnection 
\citep{2002A&ARv..10..313P, 2011LRSP....8....6S}. 
Since the buildup process of coronal free energy is 
driven by long-term evolution 
of the magnetic field on the photosphere \citep{2015ApJ...813..112T},
the features of the photospheric magnetic field, 
which can be directly observed and derived from photospheric vector magnetograms, 
may be crucial indicators for the energy transportation and triggering processes of flares/CMEs. 
These features include the size and complexity of sunspots, unsigned magnetic flux,
gradient of the magnetic field, 
magnetic energy dissipation, 
vertical electric currents, 
integrated Lorentz forces, 
magnetic shear, magnetic helicity injection, and so on
\citep{2008ApJ...686.1397P, 2009SoPh..254..101S, 2009SoPh..255...91Y, 2010AGUFMSH11B1636S}. 
With the recent development of instruments and techniques, 
it becomes easier to obtain 
extensive measurements of these features.

Many researchers have demonstrated that using
photospheric vector magnetograms in combination with
machine learning can predict solar flares effectively. 
\citet{2015ApJ...798..135B} 
described 25 features, or predictive parameters, derived from vector magnetograms 
provided by the Helioseismic and Magnetic Imager
\citep[HMI;][]{2012SoPh..275..229S} on board the 
\textit{Solar Dynamics Observatory} 
\citep[\textit {SDO};][]{2012SoPh..275....3P}.
The authors considered
flares of M1.0 class or higher, 
as defined by the peak 
1--8 \AA~flux measured by the 
\textit {Geostationary Operational Environmental Satellite} system
(\textit {GOES}). 
\citet{2017ApJ...843..104L} 
took 13 parameters out of the 25 features
and used them to perform  
multiclass predictions of solar flares.
\citet{2017ApJ...835..156N} employed both 
photospheric vector-magnetic field data and chromospheric data to predict prominent flares. 
The authors observed that 
pre-flare events such as ultraviolet brightening 
are associated with trigger mechanisms of solar flares. 
They counted the number of previous flares 
in an active region (AR) and showed that 
both the previous flare activity information 
and ultraviolet brightening are crucial for flare prediction. 
The authors later extended their study to include more features 
such as the X-ray intensity to further improve 
flare prediction performance \citep{2018ApJ...858..113N}. 
\citet{2018SoPh..293...48J} carried out solar flare prediction 
by utilizing photospheric vector-magnetic field data, 
flaring history, as well as
 multiple wavelengths of image data from the chromosphere, transition region, and corona. 
In contrast to the flare history used by 
\citet{2017ApJ...835..156N, 2018ApJ...858..113N}, \citet{2018SoPh..293...48J} 
constructed flare time series for each AR by taking
the list of associated flares in the \textit {GOES} solar-flare catalogs.
The constructed time series are then convolved with 
exponentially decaying windows of varying length
for flare prediction.

Machine learning is a subfield of artificial intelligence, 
which grants computers abilities to learn from the past data and 
make predictions on unseen future data  \citep{alpaydin2009introduction}. 
Commonly used machine learning methods for flare prediction include 
decision trees 
\citep{2009SoPh..255...91Y, 2010ApJ...709..321Y},
random forests \citep{2016ApJ...829...89B, 2017ApJ...843..104L, 2018SoPh..293...28F, breiman2001random}, 
k-nearest neighbors \citep{2008AdSpR..42.1469L, 2013A&A...549A.127H, 2015SpWea..13..286W, 2017ApJ...835..156N}, 
support vector machines \citep{2007SoPh..241..195Q, 2010RAA....10..785Y, 2015ApJ...798..135B,
2015ApJ...812...51B, 2015SpWea..13..778M, 2018SoPh..293...28F}, 
ordinal logistic regression \citep{2009SoPh..254..101S}, 
the least absolute shrinkage and selection operator (LASSO) 
\citep{2018ApJ...853...90B, 2018SoPh..293...48J}, 
extremely randomized trees \citep{2017ApJ...835..156N}, and 
neural networks \citep{2007SoPh..241..195Q, 2008AdSpR..42.1464W, 2009SpWea...7.6001C, 2011AdSpR..47.2105H, 2013SoPh..283..157A}.
Recently, \citet{2018ApJ...858..113N} adopted 
a deep neural network, named Deep Flare Net, for flare prediction.

In this paper, we attempt to use {\itshape SDO}/HMI vector magnetic field data together with flaring history 
to predict solar flares that would occur in an AR 
within 24 hours of a given time point,
with a deep learning method, named
long short-term memory 
(LSTM) \citep{DBLP:journals/neco/HochreiterS97}. 
An LSTM network is a special kind of recurrent neural networks (RNNs) \citep{Hopfield2554}
that can learn the order dependence between samples in a sequence.
LSTMs have been widely used in a variety of applications 
such as speech recognition \citep{DBLP:journals/nn/GravesS05, DBLP:conf/icann/FernandezGS07, DBLP:conf/icassp/GravesMH13}, 
handwriting recognition \citep{DBLP:conf/nips/GravesFLBS07, DBLP:conf/nips/GravesS08}, 
time series forecasting \citep{DBLP:conf/ijcai/SchmidhuberWG05} among others. 
In a solar flare prediction task, the observations in each AR form time series data, and
hence LSTMs are suitable for this prediction task. 
To our knowledge, this is the first time that 
LSTMs are used for solar flare prediction. 

The rest of this paper is organized as follows. 
Section $2$ describes our data collection scheme
and predictive parameters used in the study presented here.
Section 3 details our LSTM architecture and algorithm.
Section 4 reports experimental results.
Section 5 concludes the paper.  

\section{Data and Predictive Parameters}
\label{sec:data}
In this study we adopt the data product, named Space-weather HMI Active Region Patches \citep[SHARP;][]{2014SoPh..289.3549B}, produced by the \textit {SDO}/HMI team. 
These data were released at the end of 2012 
\citep{2014SoPh..289.3549B} 
and can be found as the \textsf{hmi.sharp} data series at 
the Joint Science Operations Center 
(JSOC).\footnote{\url{http://jsoc.stanford.edu/}}
The SHARP data encompass automatically identified and tracked ARs 
in map patches and provide many physical parameters suitable for flare prediction. 
Another useful data series, produced based on SHARP data, is \textsf{cgem.Lorentz}.
This data series includes estimations of integrated Lorentz forces 
(Sun et al. 2014)
 which help diagnose the dynamic process of each AR 
\citep{2012SoPh..277...59F}. 

Our deep learning method requires training samples.
We surveyed flares that occurred 
in the period between 2010 May and 2018 May,
using the \textit {GOES} X-ray flare 
catalogs provided by the 
National Centers for Environmental 
Information (NCEI), 
and selected flares 
with identified ARs in the NCEI flare catalogs.
This yielded a database of 
4,203 B-class flares, 
6,768 C-class flares, 
704 M-class flares, 
and 49 X-class flares. 
We used both the \textsf{hmi.sharp} data series and \textsf{cgem.Lorentz} data series, 
which were queried from the JSOC website by using SunPy \citep{2015CS&D....8a4009S}.
The data samples were collected at a cadence of 1 hour.

\begin{figure}[htbp]
	\centering\includegraphics[width=5in]{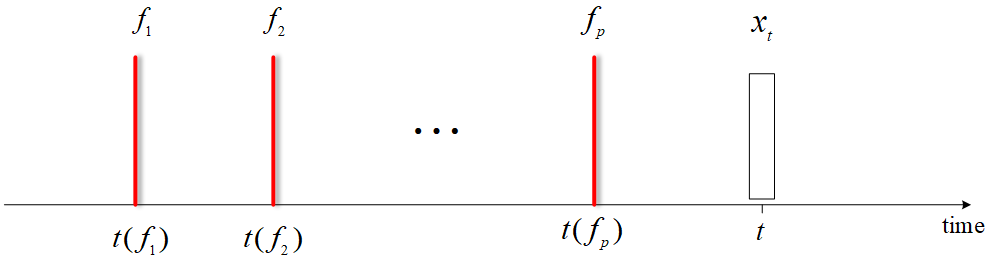} 
	\caption{Calculation of the time decay value Mdec($x_{t}$) in an AR.
		The white rectangular box represents the data sample $x_{t}$
		observed and
		collected at time point $t$ in the AR.
		There are $p$ M-class flares that occurred in the same AR prior to time point $t$, 
		so $F_{M}$ contains $p$ flares $f_{i}$, $1 \leq i \leq p$.
		Red vertical lines represent the occurrence times of the flares in $F_{M}$
		where the $i$th red vertical line indicates the starting time $t(f_i)$ 
		of the $i$th M-class flare $f_{i}$.}
	\label{fig:timedec} 
\end{figure}

We adopt two groups of predictive parameters for flare prediction.
The first group contains the 25 physical parameters
described in \citet{2015ApJ...798..135B}
that characterize AR magnetic field properties.
The second group contains 15 features related to flaring history.
Six of the 15 features are related to time decay values and
are calculated based on the formula described in 
\citet{2018SoPh..293...48J}. 
Specifically, for the data sample $x_t$ observed at time point $t$ in an AR,
the time decay value of $x_t$ with respect to 
B-class (C-class, M-class, X-class, respectively) flares, denoted
Bdec($x_{t}$) (Cdec($x_{t}$), Mdec($x_{t}$), Xdec($x_{t}$), respectively),
is computed respectively as
\begin{align}
& \text{Bdec}(x_t)=\sum_{f_i \in F_B}e^{-\frac{t-t(f_i)}{\tau}}, \\
& \text{Cdec}(x_t)=\sum_{f_i \in F_C}e^{-\frac{t-t(f_i)}{\tau}}, \\
& \text{Mdec}(x_t)=\sum_{f_i \in F_M}e^{-\frac{t-t(f_i)}{\tau}}, \\
& \text{Xdec}(x_t)=\sum_{f_i \in F_X}e^{-\frac{t-t(f_i)}{\tau}},
\end{align}
where 
$F_{B}$ ($F_{C}$, $F_{M}$, $F_{X}$, respectively) represents the set of
B-class (C-class, M-class, X-class, respectively) flares 
that occurred in the same AR before the data observation time point $t$, 
$t(f_i)$ denotes the occurrence time of flare $f_{i}$,
and $\tau$ is a decay constant that is set to 12 as suggested by \citet{2018SoPh..293...48J}. 
Figure \ref{fig:timedec} illustrates how to calculate Mdec($x_{t}$) for a data sample $x_{t}$.

The other two time decay values
for the data sample $x_t$ observed at time point $t$ in an AR
are computed by considering all
flares, regardless of their classes, that occurred in the same AR 
before the time point $t$ as follows:
\begin{equation}
	 \text{Edec}(x_t)=\sum_{f_i \in F}E(f_i) \cdot e^{-\frac{t-t(f_i)}{\tau}}, 
\end{equation}
\begin{equation}
	 \text{logEdec}(x_t)=\sum_{f_i \in F}\text{log}(E(f_i)) \cdot e^{-\frac{t-t(f_i)}{\tau}},
\end{equation}
where $F=F_B \bigcup F_C \bigcup F_M \bigcup F_X$ and $E(f_i)$ is the magnitude of flare $f_i$. 

In addition, the second group contains 9 flare history features 
for the data sample $x_t$ in the AR
as described in \citet{2017ApJ...835..156N}.
These nine features include
Bhis (Chis, Mhis, Xhis, respectively)
representing the total number of B-class 
(C-class, M-class, X-class, respectively)
flares that occurred in the same AR before 
the data observation time point $t$,
Bhis1d (Chis1d, Mhis1d, Xhis1d, respectively)
representing the total number of B-class 
(C-class, M-class, X-class, respectively)
flares that occurred in the same AR 
during the 24 hours (i.e., 1 day) prior to the time point $t$, and
Xmax1d representing the maximum X-ray intensity in the same AR 
during the 24 hours prior to the time point $t$.
In total, we use 40 features, including 25 physical features and
15 flare history features, which are summarized in  
Table \ref{tab:features}. 

\begin{table}
	\centering
	\caption{Overview of the 40 Features Including 25 \textit{SDO}/HMI Magnetic Parameters and 15 
		 Flare History Features}
	\label{tab:features}
	\begin{tabular}{lll}
	\hline
	\hline
	Keyword& Description & Formula \\ \hline
	TOTUSJH & Total unsigned current helicity &  $H_{c_{total}} \propto \sum \vert B_z \cdot J_z \vert$ \\
	TOTBSQ & Total magnitude of Lorentz force &  $F \propto \sum B^2$ \\
	TOTPOT & Total photospheric magnetic free energy density &  $\rho_{tot} \propto \sum (\pmb{B}^{\textrm{Obs}}-\pmb{B}^{\textrm{Pot}})^2dA$\\
	TOTUSJZ & Total unsigned vertical current &   $J_{z_{total}} = \sum \vert J_z \vert dA$\\
	ABSNJZH & Absolute value of the net current helicity & $H_{c_{abs}} \propto \vert \sum B_z \cdot J_z \vert$  \\
	SAVNCPP & Sum of the modulus of the net current per polarity &  $J_{z_{sum}} \propto \vert \sum^{B_z^+}J_zdA \vert + \vert \sum^{B_z^-}J_zdA \vert$ \\
	USFLUX & Total unsigned flux &  $\Phi = \sum \vert B_z \vert dA$ \\
	AREA\_ACR & Area of strong field pixels in the active region &  $\textrm{Area} = \sum \textrm{Pixels} $\\
	MEANPOT & Mean photospheric magnetic free energy &  $\overline{\rho} \propto \frac{1}{N}\sum (\pmb{B}^{\textrm{Obs}}-\pmb{B}^{\textrm{Pot}})^2$\\
	R\_VALUE & Sum of flux near polarity inversion line &  $\Phi=\sum \vert B_{LoS} \vert dA ~\textrm{within} ~R ~\textrm{mask}$ \\
	SHRGT45 & Fraction of area with shear $> 45^\circ$ &  Area with shear $>45^\circ /$ total area\\
	MEANSHR & Mean shear angle &  $\overline{\Gamma}=\frac{1}{N}\sum \textrm{arccos}(\frac{\pmb{B}^{\textrm{Obs}} \cdot \pmb{B}^{\textrm{Pot}}}{\vert \pmb{B}^{\textrm{Obs}} \vert \vert \pmb{B}^{\textrm{Pot}} \vert})$ \\
	MEANGAM & Mean angle of field from radial &  $\overline{\gamma}=\frac{1}{N}\sum \textrm{arctan}(\frac{B_h}{B_z})$\\
	MEANGBT & Mean gradient of total field &  $\overline{\vert \nabla B_{tot} \vert} = \frac{1}{N} \sum \sqrt{(\frac{\partial B}{\partial x})^2 + (\frac{\partial B}{\partial y})^2}$ \\
	MEANGBZ & Mean gradient of vertical field &  $\overline{\vert \nabla B_{z} \vert} = \frac{1}{N} \sum \sqrt{(\frac{\partial B_z}{\partial x})^2 + (\frac{\partial B_z}{\partial y})^2}$   \\
	MEANGBH & Mean gradient of horizontal field & $\overline{\vert \nabla B_{h} \vert} = \frac{1}{N} \sum \sqrt{(\frac{\partial B_h}{\partial x})^2 + (\frac{\partial B_h}{\partial y})^2}$   \\
	MEANJZH & Mean current helicity &  $\overline{H_c} \propto \frac{1}{N} \sum B_zJ_z$\\
	MEANJZD & Mean vertical current density  &  $\overline{J_z} \propto \frac{1}{N} \sum (\frac{\partial B_y}{\partial x} - \frac{\partial B_x}{\partial y})$ \\
	MEANALP & Mean characteristic twist parameter, $\alpha$ & $\alpha_{total} \propto \frac{\sum J_zB_z}{\sum B_z^2}$  \\
	TOTFX & Sum of \textit{x}-component of Lorentz force &  $F_x \propto -\sum B_xB_zdA$ \\
	TOTFY & Sum of \textit{y}-component of Lorentz force &  $F_y \propto \sum B_yB_zdA$\\
	TOTFZ & Sum of \textit{z}-component of Lorentz force &  $F_z \propto \sum (B_x^2+B_y^2-B_z^2)dA$ \\
	EPSX & Sum of \textit{x}-component of normalized Lorentz force   &  $\delta F_x \propto\frac{\sum B_xB_z}{\sum B^2}$ \\
	EPSY & Sum of \textit{y}-component of normalized Lorentz force  &  $\delta F_y \propto \frac{-\sum B_yB_z}{\sum B^2}$ \\
	EPSZ & Sum of \textit{z}-component of normalized Lorentz force  &  $\delta F_z \propto \frac{\sum(B_x^2+B_y^2-B_z^2)}{\sum B^2}$  \\
	Bdec & Time decay value based on the previous B-class flares only& $\text{Bdec}(x_t)=\sum_{f_i \in F_B}e^{-\frac{t-t(f_i)}{\tau}}$ \\
	Cdec & Time decay value based on the previous C-class flares only&  $\text{Cdec}(x_t)=\sum_{f_i \in F_C}e^{-\frac{t-t(f_i)}{\tau}}$\\
	Mdec & Time decay value based on the previous M-class flares only&  $\text{Mdec}(x_t)=\sum_{f_i \in F_M}e^{-\frac{t-t(f_i)}{\tau}}$\\
	Xdec & Time decay value based on the previous X-class flares only&  $\text{Xdec}(x_t)=\sum_{f_i \in F_X}e^{-\frac{t-t(f_i)}{\tau}}$\\
	Edec & Time decay value based on the magnitudes of all previous flares &  $\text{Edec}(x_t)=\sum_{f_i \in F}E_i \cdot e^{-\frac{t-t(f_i)}{\tau}}$\\
	logEdec & Time decay value based on the log-magnitudes of all previous flares &  $\text{logEdec}(x_t)=\sum_{f_i \in F} \text{log}(E_i) \cdot e^{-\frac{t-t(f_i)}{\tau}}$ \\
	Bhis & Total history of B-class flares in an AR & -\\
	Chis & Total history of C-class flares in an AR & -\\
	Mhis & Total history of M-class flares in an AR & -\\
	Xhis & Total history of X-class flares in an AR & -\\
	Bhis1d & 1-day history of B-class flares in an AR & -\\
	Chis1d & 1-day history of C-class flares in an AR & -\\
	Mhis1d & 1-day history of M-class flares in an AR & -\\
	Xhis1d & 1-day history of X-class flares in an AR & -\\
	Xmax1d & Maximum X-ray intensity one day before & -\\
	\hline
	\end{tabular}
\end{table}

Because the features have different units and scales,
we normalize the feature values as follows.
For the 25 physical features, 
let $z_i^k$ denote the normalized value of the $i^{th}$ feature of the $k^{th}$ data sample.
Then
\begin{equation}
z_i^k=\frac{v_i^k-\mu_i}{\sigma_i},
\label{normalization-1}
\end{equation}
where $v_i^k$ is the original value of the $i^{th}$ feature of the $k^{th}$ data sample, 
$\mu_i$ is the mean of the $i^{th}$ feature, and $\sigma_i$ is the standard deviation of the $i^{th}$ feature. 
For the 15 flare history features, we have
\begin{equation}
z_i^k=\frac{v_i^k-\text{min}_{i}}{\text{max}_{i}-\text{min}_{i}},
\label{normalization-2}
\end{equation}
where $\text{max}_{i}$ and $\text{min}_{i}$ are the maximum and minimum value of the $i^{th}$ feature, respectively.
	
\begin{table}[]
	\centering
	\caption{Numbers of Positive and Negative Samples for Each Flare Class}
	\label{tab:numSamplesEachModel}
	\begin{tabular}{c||cc||cc||cc}
		\hline
		\hline
		\multirow{2}{*}{} & \multicolumn{2}{c||}{$\ge$C class} & \multicolumn{2}{c||}{$\ge$M class} & \multicolumn{2}{c}{$\ge$M5.0 class} \\ \cline{2-7} 
		& Positive   & Negative  & Positive   & Negative  & Positive   & Negative  \\ \hline
		Training          & 18,266  &   66,311   &  2,710  &  81,867 &  633 &  83,944     \\ 
		Validation       &  7,055   &  19,418    &   1,347 &  25,126 &  292 &  26,181      \\ 
		Testing           &  8,732   &  35,957    &   1,278  & 43,411 & 180  &  44,509     \\
		\hline
	\end{tabular}
\end{table}

\section{Methodology}\label{sec:methodology}

\subsection{Prediction Task}

Following \citet{2015ApJ...798..135B}, \citet{2018SoPh..293...48J} and \citet{2018ApJ...858..113N}, 
we intend to use past observations of an AR 
to predict its future flaring activity. 
Specifically, we want to solve 
the following binary classification problem:
will this AR produce a 
$\Upsilon$-class flare within the next 24 hours?
We consider three $\Upsilon$ classes separately:
$\ge$M5.0 class, $\ge$M class, and $\ge$C class.
The importance of these classes has been discussed in recent works
\citep{2018SoPh..293...48J, 2018ApJ...858..113N}.
Both the $\ge$M class and $\ge$C class were studied in 
\citet{2018ApJ...858..113N}.
Also, the $\ge$M class was discussed in \citet{2017ApJ...843..104L}
and the $\ge$C class was analyzed in 
\citet{2018SoPh..293...48J}.
In addition, we consider the $\ge$M5.0 class due to 
the few X-class flares in our dataset
where a $\ge$M5.0-class flare means 
the \textit {GOES} X-ray flux value of the flare is above $5 \times 10^{-5}$Wm$^{-2}$. 
A flare in the $\ge$M5.0 class is generally considered a major flare. 

\begin{figure}[htbp]
	\centering\includegraphics[width=5in, trim=4 4 4 4,clip]{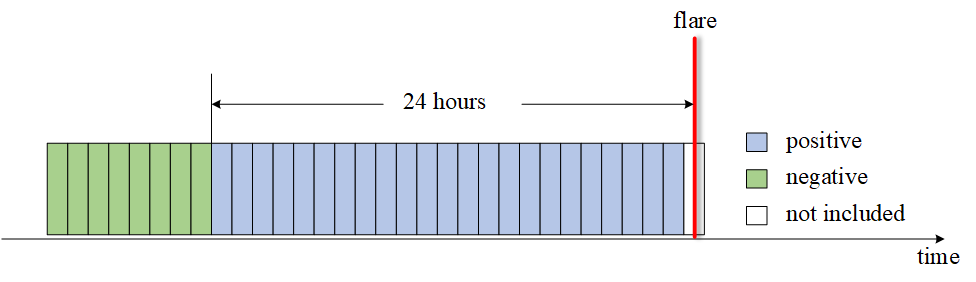} 
	\caption{
		Construction of positive and negative samples used in the prediction task.
		Each rectangular box represents a data sample, and corresponds to 1 hour in time.
		The red vertical line indicates the starting time of a $\Upsilon$-class flare. 
		Data samples collected 24 hours prior to the flare, represented by blue rectangular boxes,
		belong to the positive class.
		The other data samples, represented by green rectangular boxes, belong to the negative class.
		The white rectangular box, in which the flare occurs, is not included in our dataset.
	    }
	\label{fig:sample} 
\end{figure}

As in \citet{2015ApJ...798..135B},
observation data whose AR is outside $\pm$ $70^\circ$ of the center meridian 
or whose features are incomplete are ignored. 
Data samples collected in years 2010--2013 are used for training, 
those in year 2014 are used for validation, 
and those in years 2015--2018 are used for testing.
The training set and testing set are disjoint, and hence
our algorithm will make predictions on ARs that 
it has never seen before.
Figure \ref{fig:sample} illustrates how we 
construct positive samples and negative samples used by
the proposed deep learning method.
For the $\ge$C class, 
data samples collected 24 hours prior to an X-class, 
M-class, or C-class flare 
in an AR
are positive
and all other data samples 
in the AR
are negative.  
For the $\ge$M class, data samples collected 24 hours prior to 
an X-class or M-class flare 
in an AR
are positive
and all other data samples 
in the AR
are negative.
For the $\ge$M5.0 class, data samples collected 24 hours prior to 
a $\ge$M5.0-class flare 
in an AR
are positive and 
all other data samples 
in the AR
are negative. 
Notice that, if a data sample is missing at some time point
or if there are insufficient data samples 
during the 24 hours prior to a $\Upsilon$-class flare,
we adopt a zero-padding strategy by
adding synthetic data samples with zeros for all feature values
to yield a complete non-gapped time-series dataset.
This zero-padding method is used after applying the normalization procedures described in
Equations (\ref{normalization-1}) and (\ref{normalization-2}).
Therefore, the zero-padding method does not affect the normalization procedures.
Table \ref{tab:numSamplesEachModel} shows the numbers
of positive and negative samples for each flare class
where there are 1,269 ARs in total.
Because most ARs do not produce flares, our approach yields an
imbalanced dataset in which negative samples greatly outnumber positive samples.  

For a given time point $t$ and an AR, the proposed deep learning method
predicts whether the AR will produce a $\Upsilon$-class flare
within the next 24 hours of $t$.
There are three $\Upsilon$ classes, namely 
$\ge$M5.0 class, $\ge$M class, and $\ge$C class.
Therefore, we build three deep learning models separately,
referred to as the 
$\ge$M5.0 model, $\ge$M model, and $\ge$C model respectively, 
each corresponding to a $\Upsilon$ class of flares.

\subsection{Prediction Method}

\begin{figure}
	\centering\includegraphics[width=3.5in]{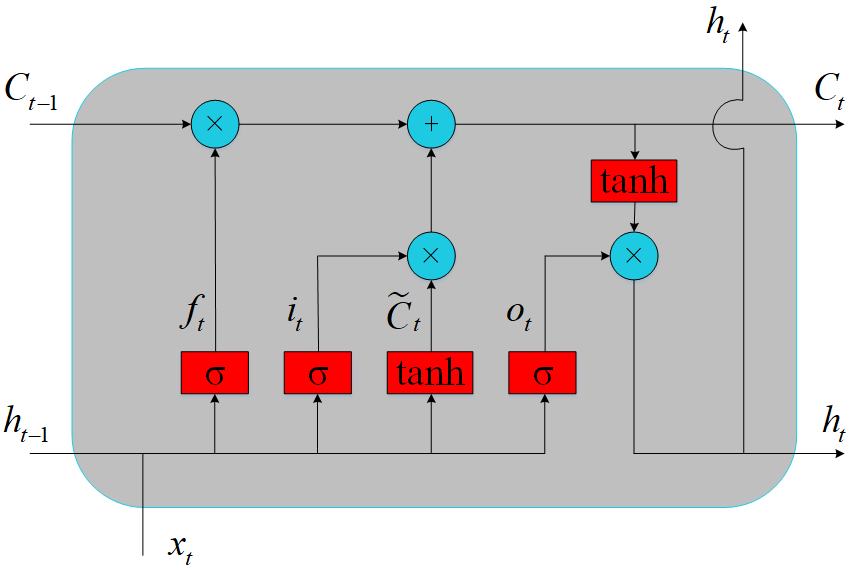} 
	\caption{Illustration of an LSTM unit. 
Here, $f_t$ is the forget gate, 
$i_t$ is the input gate,
$o_t$ is the output gate, 
 $C_t$ is the cell state,
$x_{t}$ is the input vector to the LSTM unit, and
$h_{t}$ is the output vector of the LSTM unit.
}
\label{fig:LSTM} 
\end{figure}

Our deep learning models employ 
a long short-term memory (LSTM) network. 
An LSTM unit contains four interactive parts including a memory cell, an input gate, 
an output gate and a forget gate,  as illustrated in Figure \ref{fig:LSTM}. 
The key to LSTMs is the cell state, which is represented by the horizontal line at the top of the diagram in Figure \ref{fig:LSTM}. 
Specifically, the new cell state $C_t$ is updated by the old cell state $C_{t-1}$  and the candidate cell state $\tilde{C_t}$ as follows: 
\begin{equation}
C_t = f_t \odot C_{t-1} + i_t \odot \tilde{C_t},
\end{equation}
where the forget gate $f_t$ that controls the extent to which a value remains in the cell is calculated as: 
\begin{equation}
f_t = \sigma (W_f \cdot [h_{t-1}, x_t] + B_f),
\end{equation}
and the input gate $i_t$ that controls the extent to which a new value flows into the cell is computed as:
\begin{equation}
i_t = \sigma (W_i \cdot [h_{t-1}, x_t] + B_i).
\end{equation}
Here $x_t$ represents the input vector at time step $t$ and 
$h_{t-1}$ represents the output vector at time step $t-1$.
The candidate cell state $\tilde{C_t}$ is computed as:
\begin{equation}
\tilde{C_t} = \mbox{tanh}(W_c \cdot [h_{t-1}, x_t] + B_c).
\end{equation}
Finally, the output vector $h_t$  at time step $t$, which is based on 
the new cell state $C_t$, is computed as:
\begin{equation}
h_t = o_t \odot \mbox{tanh}(C_t),
\end{equation}
where
\begin{equation}
o_t = \sigma(W_o \cdot [h_{t-1}, x_t] + B_o).
\end{equation}
In the above equations, $W$
and $B$ contain 
weights and biases respectively, which need to be learned during training;
$[.]$ denotes the concatenation of two vectors;
$\sigma(\cdot)$ is the sigmoid function, 
i.e., $\sigma(z)=\frac{1}{1+e^{-z}}$; 
$\mbox{tanh}(\cdot)$ is the hyperbolic tangent function,
i.e., $\mbox{tanh}(z)=\frac{e^z-e^{-z}}{e^z+e^{-z}}$; 
$\odot$ denotes the Hadamard product (element-wise multiplication).

\begin{figure}[htbp]
	\centering\includegraphics[width=6in]{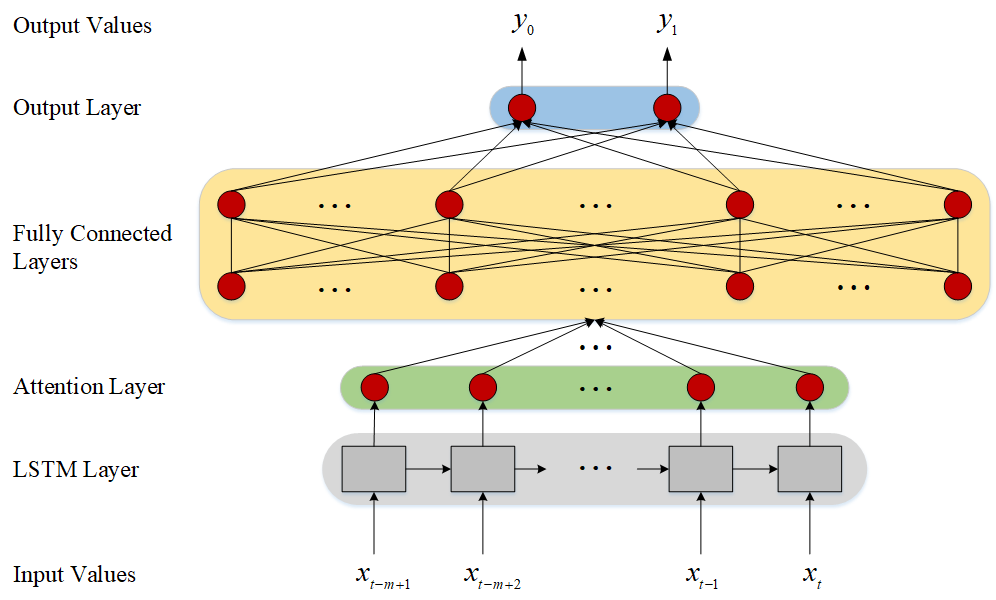} 
	\caption{Architecture of the proposed LSTM network. 
		This network is mainly comprised of an LSTM layer,
		an attention layer, two fully connected layers and an output layer.
		Each gray box in the LSTM layer is an LSTM unit as shown in Figure 3.
		There are $m$ LSTM units in the LSTM layer, 
		$m$ neurons in the attention layer, 
		200 neurons in the first fully connected layer, 500 neurons in the second fully connected layer,
		and 2 neurons in the output layer activated by the softmax function.
    	The LSTM network takes as input the sequence
    	$x_{t-m+1}, x_{t-m+2}, \dots, x_{t-1}, x_t$  
    	and produces as output a 2-dimension vector $[y_0, y_1]$ with a value of $[1, 0]$ 
	    or $[0, 1]$ which is determined by the probability 
    	calculated by the softmax function in the output layer.
}
\label{fig:architecture} 
\end{figure}

Our deep learning architecture contains an LSTM layer with $m$ LSTM units
(in the study presented here, $m$ is set to 10).
Motivated by the previous work in language translation \citep{DBLP:journals/corr/BahdanauCB14}, 
where attention mechanism was applied to allow a model to automatically search for parts 
of a source sentence that are related to the prediction of a target word, 
we add an attention layer with $m$ neurons above the LSTM layer
 to focus on information in relevant time steps. 
The attention layer would take the states in all time steps into account 
and assign a weight to each state, which indicates the importance of information that state has. 
The weight $w_i$ for state $h_i$ is derived by comparing the target state with each source state, 
which is computed as:
\begin{equation}
w_i=\frac{e^{score(h_i, h_t)}}{\sum_je^{score(h_j, h_t)}}.
\end{equation}
Here, $h_t$ is the state at the last time step and $score(\cdot)$ is a content-based function. 
We adopt the function used by \citet{DBLP:journals/corr/LuongPM15},
which is defined as 
\begin{equation}
score(h_i, h_t)=h_t^TWh_i,
\end{equation}
where $W$ contains learnable parameters. 
After $w_i$ is obtained, a context vector $c_t$ can be computed as:
\begin{equation}
c_t=\sum_iw_ih_i.
\end{equation}
The final attention vector $v$
of the input sequence 
is derived by concatenating the context vector $c_t$ and last hidden state $h_t$, 
and then being activated by a hyperbolic tangent layer as follows:
\begin{equation}
v=\textrm{tanh}(W_v[c_t;h_t]).
\end{equation}
This activation vector $v$ is then sent to 
two fully connected layers, the first one having 200 neurons and the second one having 500 neurons.
Finally the output layer with 2 neurons, which is activated by the softmax function,
produces predicted values.
Figure \ref{fig:architecture} shows the overall architecture of our LSTM network. 

Let $x_{t}$ represent the data sample collected at time point $t$.
During training, for each time point $t$,
we take $m$ consecutive data samples
 $x_{t-m+1}, x_{t-m+2}, \dots, x_{t-1}, x_t$ from the training set
 and use the $m$ consecutive data samples to train the LSTM network.
The label of these $m$ consecutive data samples is defined
to be the label of the last data sample $x_{t}$.
Thus, if $x_{t}$ belongs to the positive class, then the input sequence
 $x_{t-m+1}, x_{t-m+2}, \dots, x_{t-1}, x_t$ 
 is defined as positive; otherwise the sequence is defined as negative.
 Because the data samples are collected continuously at a cadence of 1 hour
 and missing values are filled up by our zero-padding strategy,
 the input sequence spans $m$ hours. 
 
 Because the dataset at hand is imbalanced where negative samples outnumber
 positive samples (see Table 2),
 we use a weighted cross entropy cost function for optimizing model parameters during training.
 The cost function is computed as:
 \begin{equation}\label{eq:loss}
 J=\sum_{n=1}^N\sum_{k=1}^K\omega_ky_{nk}\textrm{log}(\hat{y}_{nk}).
 \end{equation}
Here, $N$ is the total
number of sequences each having $m$ consecutive data samples
 in the training set,
$K$ is the number of classes, which is 2 in our case since we have
 only positive and negative classes, 
$\omega_k$ is the weight of the $k$th class,
 which is derived by the ratio of the sizes of the positive and negative classes
 with more weight given to the minority (i.e., positive) class, 
 $y_{nk}$ and $\hat{y}_{nk}$ denote the observed probability 
 (which is equal to $1$ if the $n$th sequence belongs to the $k$th class) 
 and the estimated probability of being in the $k$th class of the $n$th sequence, 
 respectively.\footnote{We model data samples in ARs as time series. 
However, we do not bind the many time series to a single one. 
Instead, we process ARs separately as follows.
Our LSTM network accepts as input $m$=10 data samples at a time. 
Assuming there are $P$ data samples on an active region $AR1$, 
with our zero-padding strategy, 
we generate $P$ sequences each having 10 data samples from $AR1$, 
and feed these $P$ sequences, one sequence at a time, to the LSTM network. 
Next, assuming there are $Q$ data samples on another active region $AR2$, 
we generate $Q$ sequences each having 10 data samples from $AR2$, 
and feed these $Q$ sequences, one sequence at a time, to the LSTM network. 
Although $AR1$ and $AR2$ may have overlapping time points, 
we process the active regions separately, one at a time. 
$N$ in Equation \eqref{eq:loss} represents the total number of sequences we generate from the training set.}
 
The proposed LSTM network is implemented in Python, TensorFlow and Keras. 
A mini-batch strategy \citep{DBLP:books/daglib/0040158} is used 
to achieve faster convergence during backpropagation. 
The validation dataset is used for tuning model hyperparameters.
The optimizer used is 
Adam \citep{DBLP:journals/corr/KingmaB14}, 
which is a method for stochastic gradient descent, 
where the learning rate is set to 0.001, 
$\beta_1$ is set to 0.9, and 
$\beta_2$ is set to 0.999. 
The batch size is set to 256
and the number of epochs is set to 7 .
The length of each input sequence, $m$, is set to 10, 
meaning that every time 10 consecutive data samples are used as input to our LSTM network.
 
During testing,
to predict whether an AR will produce a $\Upsilon$-class flare 
within the next 24 hours of a time point $t$,
we take $x_{t}$ and its preceding $m-1$ data samples, 
and then
feed the $m$ consecutive testing data samples
$x_{t-m+1}, x_{t-m+2}, \dots, x_{t-1}, x_t$
into the trained LSTM network.
Here, the $\Upsilon$ class refers to the
$\ge$M5.0 class, $\ge$M class, and $\ge$C class, respectively.
The output of the LSTM network, i.e., the predicted result,
 is a 2-dimension vector $[y_0, y_1]$ with a value of $[1, 0]$ or $[0, 1]$,
 indicating $x_{t}$ is positive 
 (i.e., the AR will produce a $\Upsilon$-class flare within the next 24 hours of $t$)
 or $x_{t}$ is negative
 (i.e., the AR will not produce a $\Upsilon$-class flare within the next 24 hours of $t$).
This value is determined by comparing the probability calculated by the softmax function with a threshold.
If the probability is greater than or equal to the threshold, then $x_{t}$ is predicted to be positive;
otherwise $x_{t}$ is predicted to be negative.
 It should be pointed out that, the way we use 
 the $m$ consecutive testing data samples
 $x_{t-m+1}, x_{t-m+2}, \dots, x_{t-1}, x_t$
 to predict whether there is a 
 $\ge$M5.0-class ($\ge$M-class, $\ge$C-class, respectively)
 flare within the 
 next 24 hours of the time point $t$
 is totally different from the previously published machine learning methods
 for solar flare prediction
\citep{2015ApJ...798..135B, 2018SoPh..293...48J, 2018ApJ...858..113N}, 
which used only the testing data sample $x_{t}$ to make the prediction. 

\section{Results}\label{sec:discussion}

\subsection{Performance Metrics}

Given an AR and a data sample $x_{t}$ observed at time point $t$,
we define $x_{t}$ to be a true positive (TP) if 
the $\ge$M5.0 ($\ge$M, $\ge$C, respectively) model
predicts that $x_{t}$ is positive, 
i.e., the AR will produce a 
$\ge$M5.0- ($\ge$M-, $\ge$C-, respectively) class
flare within the next 24 hours of $t$, and
$x_{t}$ is indeed positive.
We define $x_{t}$ to be a false positive (FP) if 
the $\ge$M5.0 ($\ge$M, $\ge$C, respectively) model
predicts that $x_{t}$ is positive
while $x_{t}$ is actually negative
i.e., the AR will not produce a 
$\ge$M5.0- ($\ge$M-, $\ge$C-, respectively) class
flare within the next 24 hours of $t$.
We say $x_{t}$ is a true negative (TN) if 
the $\ge$M5.0 ($\ge$M, $\ge$C, respectively) model
predicts $x_{t}$ to be negative
and $x_{t}$ is indeed negative;
$x_{t}$ is a false negative (FN) if 
the $\ge$M5.0 ($\ge$M, $\ge$C, respectively) model
predicts $x_{t}$ to be negative
while $x_{t}$ is actually positive.
We also use TP (FP, TN, FN, respectively) to represent the total number of 
true positives (false positives, true negatives, false negatives, respectively)
produced by a model.

The performance metrics used in this study include the following:
\begin{equation}
 \text{Recall} = \frac{\mbox{TP}}{\mbox{TP + FN}}, 
\end{equation}
\begin{equation}
 \text{Precision} = \frac{\mbox{TP}}{\mbox{TP + FP}} , 
 \end{equation}
 \begin{equation}
\text{Accuracy (ACC)} = \frac{\mbox{TP + TN}}{\mbox{TP + FP + TN + FN}}, 
\end{equation}
 \begin{equation}
 \text{Balanced Accuracy (BACC)} = 
 \frac{1}{2}(\frac{\mbox{TP}}{\mbox{TP + FN}} + \frac{\mbox{TN}}{\mbox{TN + FP}}),  
 \end{equation}
 \begin{equation}
\text{Heidke Skill Score (HSS)} = \frac{2(\mbox{TP}\times \mbox{TN}-\mbox{FP}\times \mbox{FN})}{(\mbox{TP+FN})(\mbox{FN+TN})+(\mbox{TP+FP})(\mbox{FP+TN})},
\end{equation}
  \begin{equation}
 \text{True Skill Statistics (TSS)} = \frac{\mbox{TP}}{\mbox{TP + FN}} - \frac{\mbox{FP}}{\mbox{TN + FP}}.
  \end{equation}

ACC is not suitable for imbalanced classification \citep{DBLP:journals/tkde/HeG09}.
The reason is that a naive classifier predicting all instances 
in the minority class to belong to the majority class
would still get a high ACC value.
Instead, BACC is suggested for imbalanced classification \citep{DBLP:journals/tkde/HeG09}.
Because of its unbiasedness over class-imbalance ratios, 
we also follow the suggestion of  \citet{2012ApJ...747L..41B}
to use the TSS score, which is the recall subtracted by the false alarm rate. 
The Heidke Skill Score (HSS)
\citep{heidke1926berechnung} is used to measure the fractional improvement of our prediction over the random prediction
\citep{2018SoPh..293...28F}.
The larger BACC, HSS, and TSS score a method has,
the better performance the method achieves.

\subsection{Model Evaluation}

We first conduct an ablation study to analyze the proposed LSTM framework by
considering three alternative architectures, 
denoted by LSTM$_{-a}$, LSTM$_{-c}$ and LSTM$_{-ac}$, respectively.
LSTM$_{-a}$ (LSTM$_{-c}$, LSTM$_{-ac}$, respectively) 
is obtained by removing the attention layer 
(the two fully connected layers, 
both the attention layer 
and the two fully connected layers, respectively)
from the LSTM architecture in Figure 4.
Table \ref{tab:ablation} shows the prediction results of the four architectures
for the three $\Upsilon$ classes, namely
$\ge$M5.0 class, $\ge$M class, and $\ge$C class, of flares. 
It can be seen from Table \ref{tab:ablation} that 
the proposed LSTM architecture outperforms the ablations LSTM$_{-a}$, LSTM$_{-c}$ and LSTM$_{-ac}$
in terms of BACC, HSS and TSS scores. 
By including the attention layer and the two fully connected layers, 
the performance of the LSTM framework improves.
Our zero-padding strategy does not negatively affect the prediction performance.  
Although {\em some} normalized feature values of a data sample may become zeros due to the
 normalization procedures described in
Equations (\ref{normalization-1}) and (\ref{normalization-2}),
the attention layer of our LSTM model is able to distinguish between this data sample 
and those synthetic data samples added by our zero-padding method whose feature values are {\em all} zeros.
The attention layer pays little attention to the synthetic data samples whose feature values are all zeros.

\begin{table}[h]
	\centering
	\caption{Flare Prediction Results (within 24 hours) of Four LSTM Architectures}
	\label{tab:ablation}
	\begin{tabular}{cl||c||c||c}
		\hline
		&              & $\ge$M5.0 class & $\ge$M class & $\ge$C class \\ \hline
		\multirow{4}{*}{Recall} 
		& LSTM$_{-ac}$  & 0.944   & 0.888  & 0.743 \\
		& LSTM$_{-a}$  &  0.939 & 0.899 & 0.747 \\
	    & LSTM$_{-c}$  & 0.956  & 0.876  & 0.750     \\
		& LSTM & \textbf{0.978}  & \textbf{0.881} & \textbf{0.762}   \\ \hline
		
		{\multirow{4}{*}{Precision}} 
		& LSTM$_{-ac}$ & \textbf{0.042}   & 0.181   & 0.543   \\
		& LSTM$_{-a}$  &  0.039 & 0.184 & 0.537\\
		& LSTM$_{-c}$ & 0.041  & 0.216  & 0.536       \\
		& LSTM  & 0.038  & \textbf{0.222}  & \textbf{0.544}  \\ \hline
		
		\multirow{4}{*}{ACC} 
		& LSTM$_{-ac}$ & \textbf{0.914} & 0.882 & 0.828    \\
		& LSTM$_{-a}$  & 0.904  & 0.883 & 0.825\\
		& LSTM$_{-c}$ & 0.910   & 0.906  & 0.824  \\
		& LSTM  & 0.899  & \textbf{0.909}  & \textbf{0.829}  \\ \hline
		
		\multirow{4}{*}{BACC}  
		& LSTM$_{-ac}$ & 0.929 & 0.885  & 0.796 \\
		& LSTM$_{-a}$  & 0.933  & 0.891 & 0.795 \\
		& LSTM$_{-c}$ & 0.933  & 0.891    & 0.796 \\
		& LSTM &  \textbf{0.938}   & \textbf{0.895} & \textbf{0.803}   \\ \hline
		
		\multirow{4}{*}{HSS}  
		& LSTM$_{-ac}$ & 0.074 & 0.267  & 0.519 \\
		& LSTM$_{-a}$  & 0.068  & 0.271 & 0.515 \\
		& LSTM$_{-c}$ & 0.071  & 0.316    & 0.514 \\
		& LSTM &  \textbf{0.074}   & \textbf{0.347} & \textbf{0.539}   \\ \hline
		
		\multirow{4}{*}{TSS} 
		& LSTM$_{-ac}$      &  0.858    & 0.770  & 0.591 \\
		& LSTM$_{-a}$  &  0.865 & 0.782 & 0.591\\
		& LSTM$_{-c}$    & 0.865   & 0.783  & 0.592   \\
		& LSTM  & \textbf{0.877} & \textbf{0.790} & \textbf{0.607}   \\ \hline
	\end{tabular}
\end{table}

We next compare our LSTM framework 
with five closely related machine learning methods including
 multilayer perceptrons (MLP) \citep{haykin2004comprehensive, 2018SoPh..293...28F},
Jordan network (JN) \citep{jordan1997serial},
 support vector machines (SVM) \citep{2007SoPh..241..195Q, 2010RAA....10..785Y, 2015ApJ...798..135B, 2015ApJ...812...51B, 2015SpWea..13..778M, 2018SoPh..293...28F},
random forests (RF) \citep{2016ApJ...829...89B, 2017ApJ...843..104L, 2018SoPh..293...28F},
and a recently published deep learning-based method, Deep Flare Net \citep[DeFN;][]{2018ApJ...858..113N}.
All these methods including ours (LSTM) can be used as 
a binary classification model \citep{2018ApJ...858..113N, 2018SoPh..293...48J}
or a probabilistic forecasting model \citep{2018SoPh..293...28F}.
A binary classification model
predicts whether or not an AR will produce a
$\ge$M5.0- ($\ge$M-, $\ge$C-, respectively)
 class flare within the next 24 hours. 
A probabilistic forecasting model predicts the probability for an AR to produce a 
$\ge$M5.0- ($\ge$M-, $\ge$C-, respectively)
class flare within the next 24 hours.
A probabilistic forecasting model can be converted into a binary classification model
by using a probability threshold to make predictions as follows.
If the predicted probability is greater than or equal to the threshold, then the AR will produce a flare within the next 24 hours;
otherwise the AR will not produce a flare within the next 24 hours.

The MLP method consists of
an input layer, an output layer and two hidden layers with 200 neurons and 500 neurons respectively. 
For the JN method, its output dimension is set to 10.
SVM uses the radial basis function (RBF) kernel.
RF has two parameters: mtry (number of features randomly selected to split a node)
and ntree (number of trees to grow in the forest).
We vary the values of ntree $\in$ \{300, 500, 1,000\} and mtry $\in$ [2, 8],
and set ntree to 500 and mtry to 3 since these two parameter values yield the maximum TSS for RF.
Table \ref{tab:comparison} compares our LSTM with the five related methods.
All the methods are treated as binary classification models where
the probability thresholds are chosen to maximize their respective TSS values, and
the same threshold is used to calculate all performance metrics
including BACC, HSS and TSS for each method with respect to each flare class.
It can be seen from 
Table \ref{tab:comparison} that LSTM and RF are the two best methods.
The probability thresholds used by LSTM in Table \ref{tab:comparison} are 
50\%, 60\% and 50\% for $\ge$M5.0, $\ge$M and $\ge$C models respectively
(the same thresholds are used in Table \ref{tab:ablation}).
The probability thresholds used by RF in Table \ref{tab:comparison} are 
0.5\%, 5\% and 25\% for $\ge$M5.0, $\ge$M and $\ge$C models respectively. 

\begin{table}[h]
	\centering
	\caption{Flare Prediction Results (within 24 hours) of Our LSTM and Five Related Machine Learning Methods}
	\label{tab:comparison}
	\begin{tabular}{cc||c||c||c}
		\hline
		&              & $\ge$M5.0 class & $\ge$M class & $\ge$C class \\ \hline
		\multirow{6}{*}{Recall} & MLP  & 0.944   & 0.812  & 0.637 \\
		& SVM  & 0.644   & 0.692  & 0.746 \\
		& JN  & 0.923   & 0.851  & 0.701 \\
		& DeFN         & 0.889   & \textbf{0.891}   & 0.761   \\
		& RF  & \textbf{1.000}  & 0.850  & 0.727     \\
		& LSTM & 0.978  & 0.881 & \textbf{0.762}   \\ \hline
		
		{\multirow{6}{*}{Precision}} & MLP  & 0.037   & 0.143  & 0.451 \\
		& SVM & 0.014   & 0.106   & 0.497   \\
		& JN     & 0.033   & 0.178  & 0.543 \\
		& DeFN & 0.037   & 0.173 & 0.497   \\
		& RF & 0.034  & \textbf{0.252}  & 0.532       \\
		& LSTM  & \textbf{0.038}  & 0.222  & \textbf{0.544}  \\ \hline
		
		\multirow{6}{*}{ACC} & MLP  & 0.901   & 0.855  & 0.778 \\
		& SVM & 0.818   & 0.824 & 0.803    \\
		& JN     & 0.882   & 0.884  & 0.826 \\
		& DeFN         & \textbf{0.907}  & 0.872 & 0.801   \\
		& RF & 0.886   & \textbf{0.924}  & 0.822  \\
		& LSTM  & 0.899  & 0.909  & \textbf{0.829} \\ \hline
		
		\multirow{6}{*}{BACC}    & MLP  & 0.922   & 0.834  & 0.725 \\
		& SVM & 0.732 & 0.760  & 0.781 \\
		& JN   & 0.903   & 0.868  & 0.779 \\
		& DeFN         & 0.898   & 0.881 & 0.786 \\
		& RF & \textbf{0.943}  & 0.888    & 0.786 \\
		& LSTM &  0.938   & \textbf{0.895} & \textbf{0.803}   \\ \hline
		
		\multirow{6}{*}{HSS}    & MLP  & 0.064   & 0.204  & 0.389 \\
		& SVM & 0.020 & 0.141  & 0.472 \\
		& JN    & 0.056   & 0.260  & 0.502 \\
		& DeFN         & 0.064   & 0.253 & 0.476 \\
		& RF & 0.059  & \textbf{0.361}    & 0.502 \\
		& LSTM &  \textbf{0.074}   & 0.347 & \textbf{0.539}   \\ \hline
		
		\multirow{6}{*}{TSS} & MLP  & 0.845   & 0.669  & 0.449 \\
		& SVM           &  0.464    & 0.520  & 0.562 \\
		& JN             & 0.806   & 0.736  & 0.558 \\
		& DeFN         & 0.796  & 0.763 & 0.572 \\
		& RF           & \textbf{0.886}   & 0.776  & 0.572   \\
		& LSTM  & 0.877 & \textbf{0.790} & \textbf{0.607}   \\ \hline
	\end{tabular}
\end{table}

To further understand the behavior of the proposed LSTM method, we conduct a cross-validation study as follows. 
Refer to Table 2.
For each of the $\ge$M5.0, $\ge$M and $\ge$C models, 
we partition its training (testing, respectively) set into 10 equal-sized folds.
For every two training (testing, respectively) folds $ i$ and $ j$, $i \not= j$, fold $ i$ and fold $ j$ are disjoint; 
furthermore, fold $i$ and fold $j$ contain approximately the same number of positive training (testing, respectively) data samples
and approximately the same number of negative training (testing, respectively) data samples. 
In run ($i$, $j$), $1 \leq i \leq 10$, $1 \leq j \leq 10$,
all training samples except those in training fold $i$ are used to train a model,
and the trained model is used to make predictions on all testing samples except those in testing fold $j$.
We calculate the performance metric values based on the predictions made in run ($i$, $j$).
There are 100 runs.
The means and standard deviations over the 100 runs are calculated and recorded.

\subsection{Feature Assessment}

Motivated by RF which uses only 3 features
to split a node when constructing a tree,
we wonder whether using fewer features can also achieve better performance for our LSTM method.
We thus analyze the importance of each of the 40 features studied in the paper
with respect to the 
$\ge$M5.0, $\ge$M, and $\ge$C models respectively
based on the LSTM architecture in Figure 4
using the cross-validation methodology described above.
Each time only one feature is used by the 
$\ge$M5.0 ($\ge$M, $\ge$C, respectively) model
to predict whether a given AR will produce a 
$\ge$M5.0- ($\ge$M-, $\ge$C-, respectively) class flare
within the next 24 hours.
The probability threshold is set to maximize the TSS score in each test.
The corresponding mean TSS score is recorded.
There are 40 features, so 40 mean individual TSS scores are recorded.
These 40 mean individual TSS scores are sorted in descending order, and the 40 corresponding features are ranked
from the most important to the least important accordingly.
The sorted, mean individual TSS scores and ranked features 
are plotted in a chart for each model.
Then, according to the ranked features, mean cumulative TSS scores are calculated
and plotted in the chart. 
Specifically, the mean cumulative TSS score of the top $k$,
$1 \leq k \leq 40$,
most important features
with respect to the 
$\ge$M5.0 ($\ge$M, $\ge$C, respectively) model
is equal to the mean TSS score of the 
$\ge$M5.0 ($\ge$M, $\ge$C, respectively) model
that uses the top $k$ most important features altogether for flare prediction. 

\begin{figure}[htbp]
	\centering
	\includegraphics[width=6.1in, trim=4 2 4 2,clip]{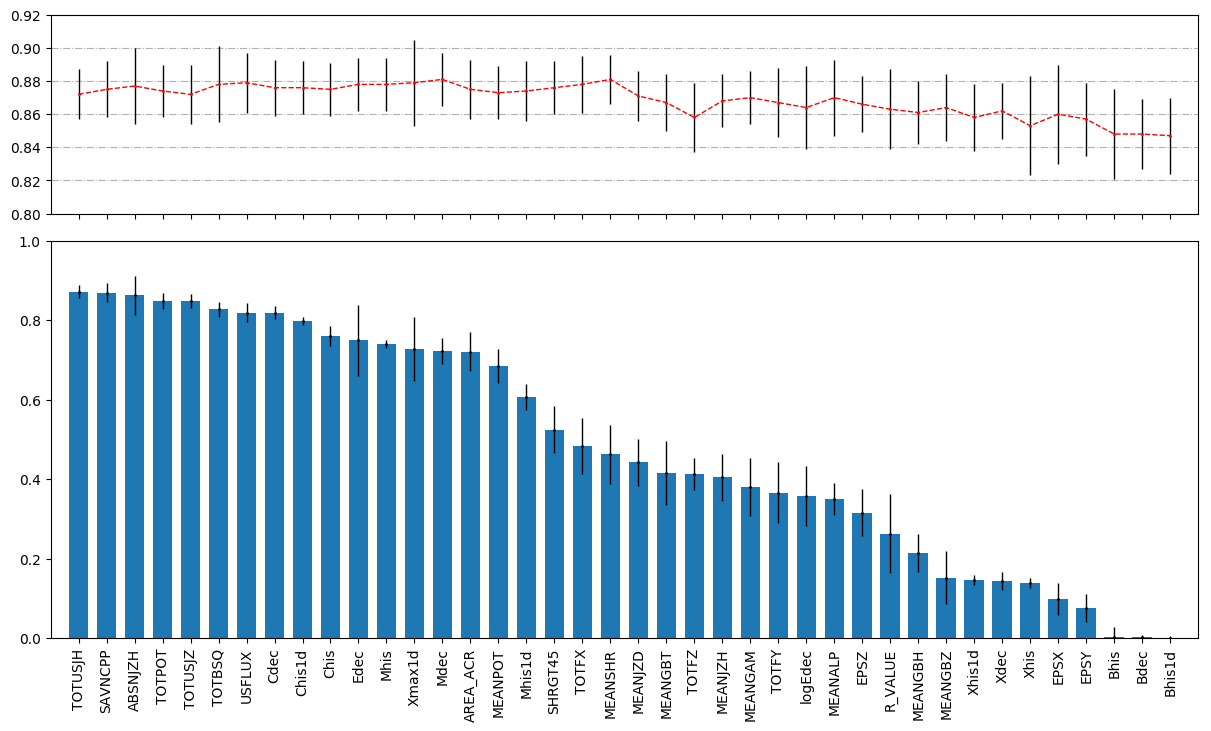}  \\
	\caption{Assessment of feature importance for predicting $\ge$M5.0-class flares.
		Blue bars represent mean individual TSS scores and
		the red polygonal line represents mean cumulative TSS scores of the features.}
	\label{fig:M5.0importance} 
\end{figure}

\begin{figure}[htbp]
	\centering
	\includegraphics[width=6.1in, trim=4 2 4 2,clip]{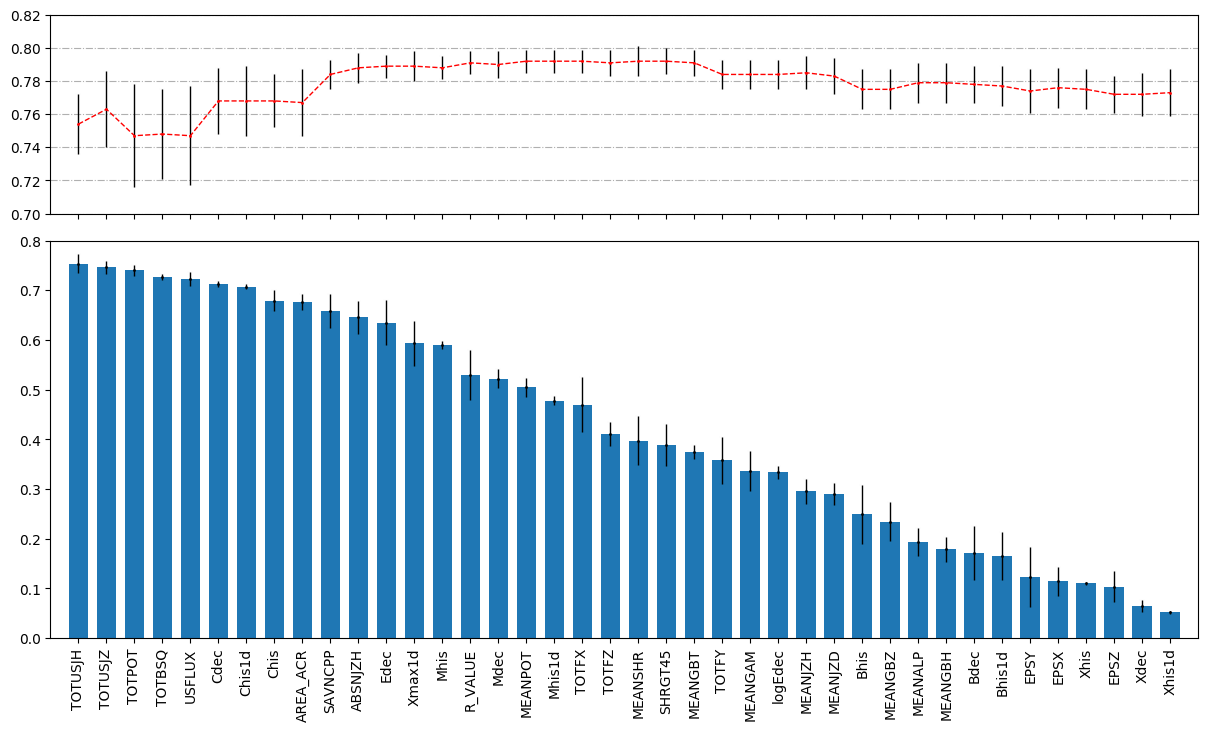}   \\
	\caption{
		Assessment of feature importance for predicting $\ge$M-class flares.
		Blue bars represent mean individual TSS scores and
		the red polygonal line represents mean cumulative TSS scores of the features.}
	\label{fig:Mimportance} 
\end{figure}

\begin{figure}[htbp]
	\centering
	\includegraphics[width=6.1in, trim=4 2 4 2,clip]{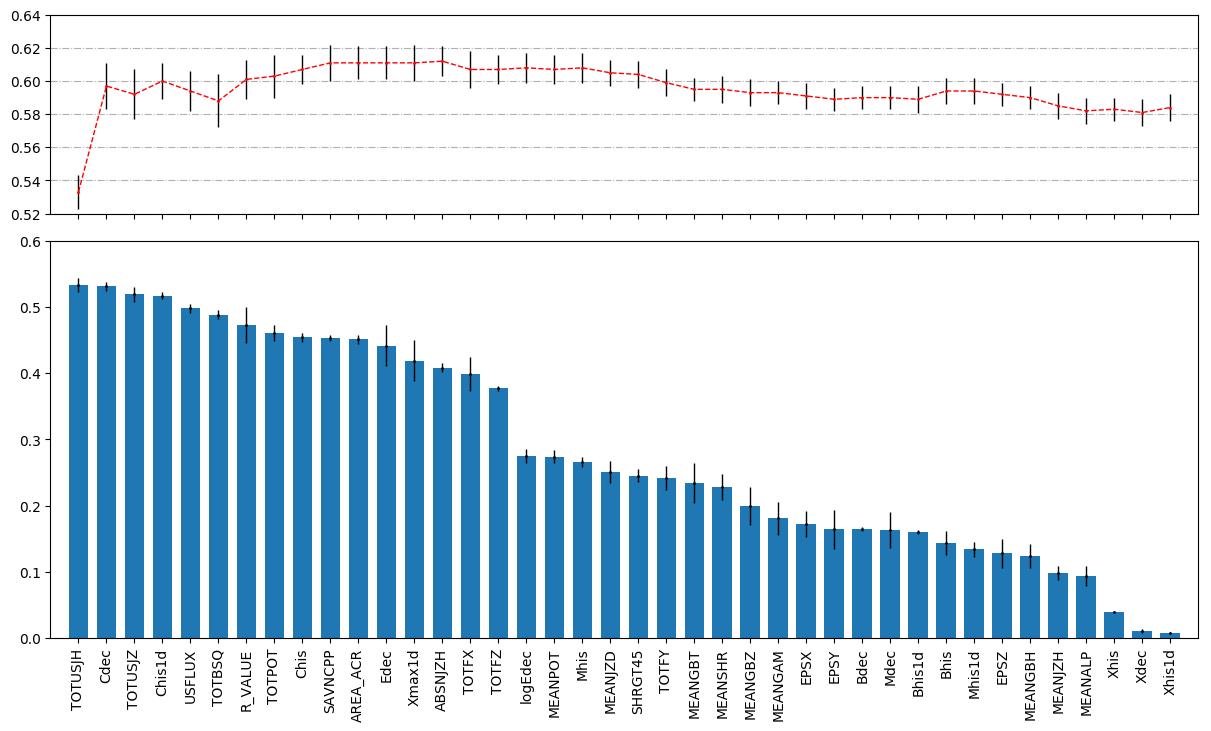} 
	\vspace{0.1cm}
	\caption{
		Assessment of feature importance for predicting $\ge$C-class flares.
		Blue bars represent mean individual TSS scores and
		the red polygonal line represents mean cumulative TSS scores of the features.}
	\label{fig:Cimportance} 
\end{figure}

Figure \ref{fig:M5.0importance} 
(\ref{fig:Mimportance}, \ref{fig:Cimportance}, respectively)
presents the feature importance chart for the
$\ge$M5.0 ($\ge$M, $\ge$C, respectively) model. 
In each figure, blue bars represent mean individual TSS scores of the 40 features
and the red polygonal line represents mean cumulative TSS scores
of the top $k$,
$1 \leq k \leq 40$,
most important features.
Error bars, representing standard deviations, are also plotted.
It can be seen from the figures that predictive parameters that
are consistently ranked in the top 20 list for all the three models
include the following 16 features: 
TOTUSJH, TOTBSQ, TOTPOT, TOTUSJZ, ABSNJZH, SAVNCPP, USFLUX, AREA\_ACR, 
MEANPOT, TOTFX, 
Cdec, Chis, Chis1d, Edec, Mhis and Xmax1d. 
Among these 16 features, there are 10 physical parameters, or
\textit{SDO}/HMI magnetic parameters, including
TOTUSJH, TOTBSQ, TOTPOT, TOTUSJZ, ABSNJZH, SAVNCPP, USFLUX, AREA\_ACR, 
MEANPOT and TOTFX.
All these 10 physical parameters except TOTFX are among the 13 magnetic parameters that are also considered important in 
\citet{2015ApJ...798..135B} and \citet{2017ApJ...843..104L}, which used
different methods for assessing the importance of features.
Thus, our findings are consistent with those reported in the literature.
There are 4 magnetic parameters, 
TOTFZ, R\_VALUE, SHRGT45 and EPSZ, which are considered important in
\citet{2015ApJ...798..135B} and \citet{2017ApJ...843..104L},
but are not ranked high in our list;
some flare history features are ranked higher than these four magnetic parameters. 
It is worth noting that TOTUSJH plays the most important role, 
i.e., is ranked the top one, for all the three models. 
Moreover, some features including 
TOTUSJH, TOTUSJZ, TOTPOT, TOTBSQ, USFLUX, SAVNCPP, Cdec, Chis and Chis1d
show strong predictive power
and are consistently ranked in the top 10 list for all the three models.

We note that the history of C-class flares 
has a high impact on flare prediction.
Chis and Chis1d from \citet{2017ApJ...835..156N} count the number of previous C-class flares.
Cdec from \citet{2018SoPh..293...48J} is the time decay value based on previous C-class flares.
These three flare history features are ranked high in our list.
The history of M-class flares plays a more important role for the $\geq$M5.0 
and $\geq$M models than for the $\geq$C model.
Other flare history features such as the histories of B-class and X-class flares
are relatively unimportant for predicting 
$\geq$M5.0- ($\geq$M-, $\geq$C-, respectively) class flares.

By carefully examining Figures \ref{fig:M5.0importance}, 
\ref{fig:Mimportance} and \ref{fig:Cimportance},
we find that using all the 40 features together does not 
yield the highest mean cumulative TSS scores.
In fact, using roughly the top 14-22 most important features together
yields the highest mean cumulative TSS scores,
achieving the best performance.
Specifically, using the top 20 (22, 14, respectively) most important features
yields the highest mean cumulative TSS score for the $\geq$M5.0
($\geq$M, $\geq$C, respectively) model.
 This happens probably because low ranked features are noisy features, and
using them may deteriorate the performance of the models.
In subsequent experiments, we use the best features for each model.

\subsection{Comparison between RF and LSTM}

Table \ref{tab:comparison} shows that
RF and LSTM are the two best methods. 
In this subsection, we further compare RF and LSTM using the cross-validation methodology described above.
Table \ref{tab:twobest} shows their performance metric values where 
standard deviations are enclosed in parentheses.
The probability thresholds used by RF 
are set to 
0.5\%, 5\%, 25\% for $\ge$M5.0, $\ge$M, $\ge$C class respectively
where the thresholds are chosen to maximize TSS.
The probability thresholds used by LSTM are set to
75\%, 60\%, 50\% for $\ge$M5.0, $\ge$M, $\ge$C class respectively.
These optimal thresholds are slightly different from those used in Table \ref{tab:comparison}. 
This happens because the performance metric values in Table \ref{tab:comparison} are obtained from a single dataset
whereas those in Table \ref{tab:twobest} are obtained from cross validation.
Furthermore, LSTM in Table \ref{tab:comparison} uses all 40 features together
whereas LSTM in Table \ref{tab:twobest} uses only the (14-22) best features.
According to Table \ref{tab:twobest} and a Wilcoxon signed rank test \citep{Wilcoxon-1947}, 
our LSTM method is significantly better than RF (p $<$ 0.05)
in all three flare classes in terms of BACC, HSS and TSS.\footnote{The
source code of LSTM and dataset can be downloaded from
\url{https://web.njit.edu/~wangj/LSTMpredict/}.}
These results indicate that LSTM outperforms RF when the methods are used as binary classification models.

\begin{table}
	\centering
	\caption{Flare Prediction Results (within 24 hours) of Our LSTM and RF Obtained from Cross Validation}
	\label{tab:twobest}
	\begin{tabular}{cc||c||c||c}
		\hline
		&              & $\ge$M5.0 class & $\ge$M class & $\ge$C class \\ \hline
		\multirow{2}{*}{Recall} & RF  & 0.960 (0.027)   & \textbf{0.888} (0.006)  & 0.730 (0.002) \\
		& LSTM & \textbf{0.960} (0.017)  & 0.885 (0.017) & \textbf{0.773} (0.027)  \\ \hline
		
		{\multirow{2}{*}{Precision}} & RF & 0.026 (0.001)   & 0.179 (0.003)   & 0.499 (0.003)   \\
		& LSTM & \textbf{0.048} (0.008)  & \textbf{0.222} (0.023)  & \textbf{0.541} (0.030)       \\ \hline
		
		\multirow{2}{*}{ACC} & RF & 0.853 (0.003)   & 0.880 (0.002)  & 0.804 (0.001)    \\
		& LSTM & \textbf{0.921} (0.014)   & \textbf{0.907} (0.013)  & \textbf{0.826} (0.015)  \\ \hline
		
		\multirow{2}{*}{BACC}    & RF & 0.906 (0.014) & 0.884 (0.003)  & 0.776 (0.001) \\
		& LSTM & \textbf{0.940} (0.007)  & \textbf{0.896} (0.004)    & \textbf{0.806} (0.004) \\  \hline
		
		\multirow{2}{*}{HSS}    & RF & 0.042 (0.002) & 0.262 (0.004)  & 0.469 (0.003) \\
		& LSTM & \textbf{0.084} (0.015)  & \textbf{0.323} (0.030)    & \textbf{0.526} (0.021) \\  \hline
		
		\multirow{2}{*}{TSS} & RF           &  0.812 (0.027)    & 0.768 (0.005)  & 0.552 (0.003) \\
		& LSTM           & \textbf{0.881}  (0.014)   & \textbf{0.792}  (0.008)  & \textbf{0.612}  (0.009)  \\ \hline
	\end{tabular}
\end{table}

We next compare RF and LSTM using 
(i)  skill scores profiles (SSP) of BACC, HSS and TSS as functions of the probability threshold,
(ii) Receiver Operating Characteristic (ROC) curves, and
(iii) Reliability Diagrams (RD).
ROC curves describe the relationship between the true positive rate and false positive rate. 
The Area Under the Curve (AUC) in the ROC represents the degree of separability, indicating 
how well a model is capable of distinguishing between two classes with the ideal value of one \citep{marzban2004roc}. 
The RD describes the relationship between the actual observed frequencies of flares of interest and the probabilities predicted by a model. 
A bin diagram, presented as an inset in the RD, is used to show the distribution of the predicted probabilities.
The X-axis of the bin diagram represents the predicted probabilities and 
the Y-axis represents the numbers of data samples.
As in \citet{2018SoPh..293...28F} we use 20 bins of length 0.05 each.
Thus, for example, the Y value of the first bin shows the number of data samples
whose predicted probabilities of having a flare of interest in the next 24 hours are within 0.05.
The ideal situation is represented by the diagonal line in the RD;
a point $(x, x)$ on the diagonal indicates that
among those data samples whose predicted probability is $x$,
the ratio of the data samples that actually have a flare of interest within the next 24 hours is also $x$.
In addition, we use the Brier Score (BS) \citep{brier1950verification} 
and Brier Skill Score (BSS) \citep{wilks2010sampling, wilks2011statistical}
to quantitatively assess the performance of probabilistic forecasting models.
The values of BS range from 0 to 1 with the perfect score being 0.
The values of BSS range from minus infinity to 1 with the perfect score being 1.

\begin{figure}
	\gridline{\fig{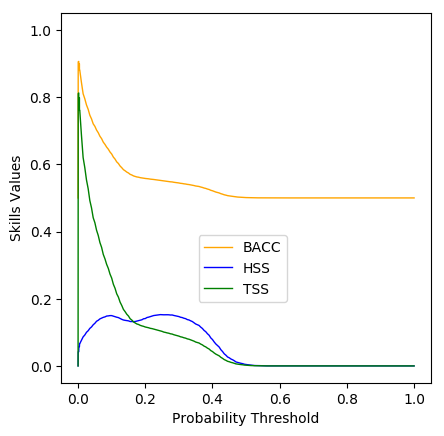}{0.33\textwidth}{(a) RF, SSP}
		\fig{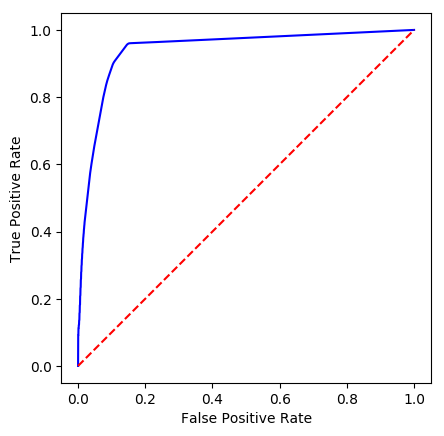}{0.33\textwidth}{(b) RF, ROC}
		\fig{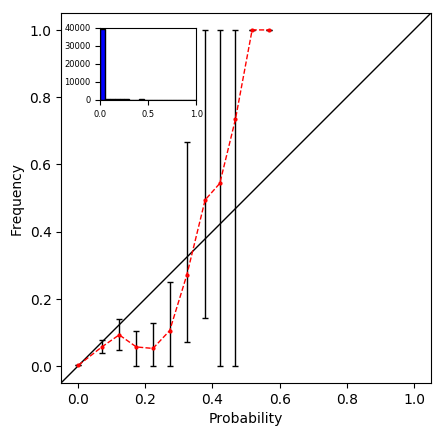}{0.33\textwidth}{(c) RF, RD}}
	\gridline{\fig{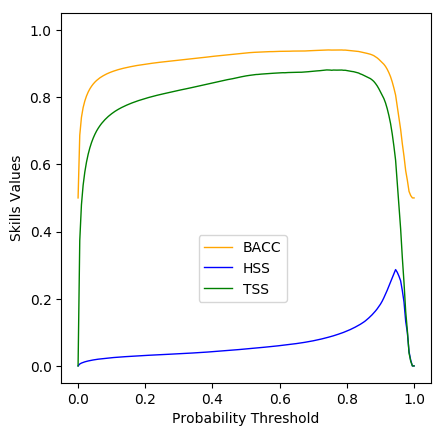}{0.33\textwidth}{(d) LSTM, SSP}
		\fig{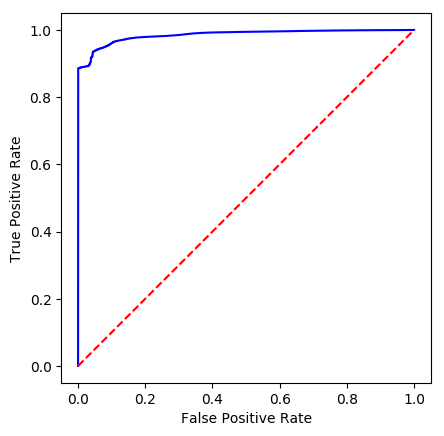}{0.33\textwidth}{(e) LSTM, ROC}
		\fig{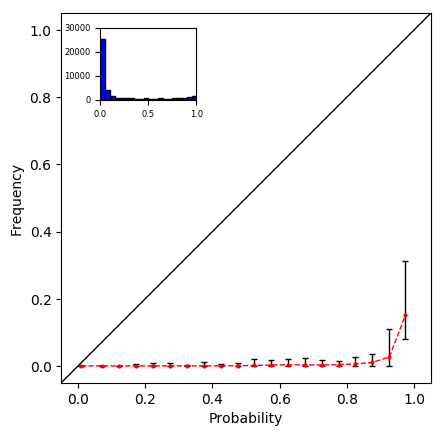}{0.33\textwidth}{(f) LSTM, RD}}
	\caption{Comparison between RF (top) and LSTM (bottom) for $\geq$M5.0-class flare prediction
where the corresponding SSP, ROC and RD are displayed from left to right.}
	\label{fig:M5visual} 
\end{figure}

Figure \ref{fig:M5visual} presents SSP, ROC and RD plots 
for RF and LSTM respectively
when the methods are used in
$\geq$M5.0-class flare prediction.
Refer to the SSP plots.
For RF, the maximum TSS=0.812 is obtained with a probability threshold of 0.5\%.
With this threshold, BACC=0.906$\pm$0.014 and HSS=0.042$\pm$0.002.
For LSTM, the maximum TSS=0.881 is obtained with a probability threshold of 75\%.
With this threshold, BACC=0.940$\pm$0.007 and HSS=0.084$\pm$0.015.
The ROC curve of LSTM is better than that of RF. 
LSTM has an AUC of 0.984$\pm$0.003, which is better than RF with an AUC of 0.948$\pm$0.011. 
Refer to the RD plots. 
The curves of RF and LSTM are far away from the diagonal lines in the RD plots. 
The BS and BSS achieved by RF are 0.004$\pm$0.001 and 0.053$\pm$0.008 respectively. 
The BS and BSS achieved by LSTM are 0.090$\pm$0.011 and -21.576$\pm$2.956 respectively.
In terms of BS and BSS, RF is better than LSTM.

\begin{figure}
	\gridline{\fig{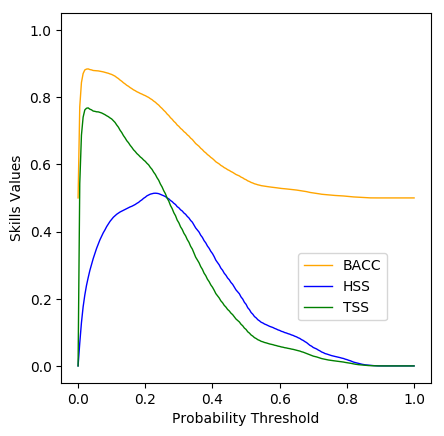}{0.33\textwidth}{(a) RF, SSP}
		\fig{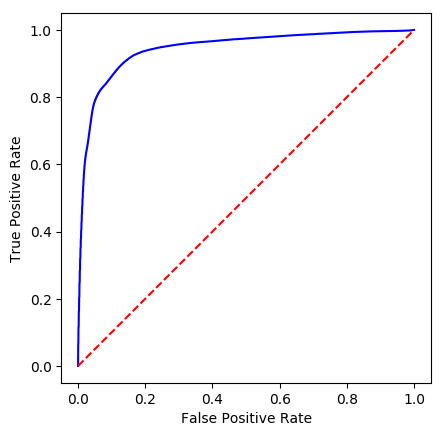}{0.33\textwidth}{(b) RF, ROC}
		\fig{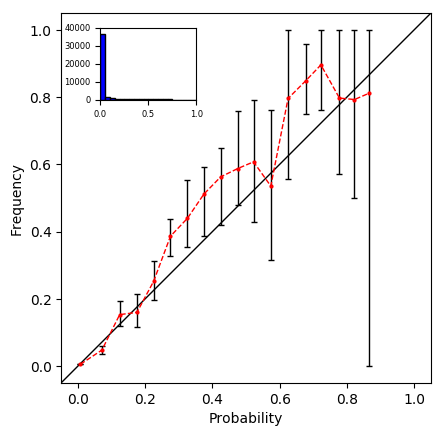}{0.33\textwidth}{(c) RF, RD}}
	\gridline{\fig{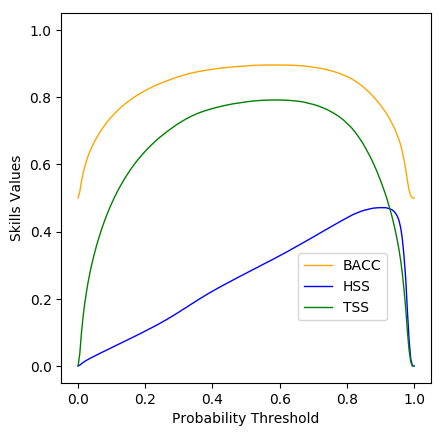}{0.33\textwidth}{(d) LSTM, SSP}
		\fig{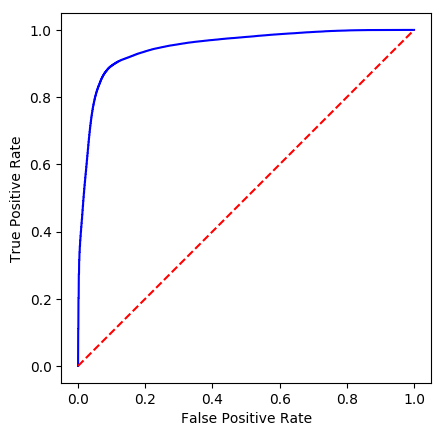}{0.33\textwidth}{(e) LSTM, ROC}
		\fig{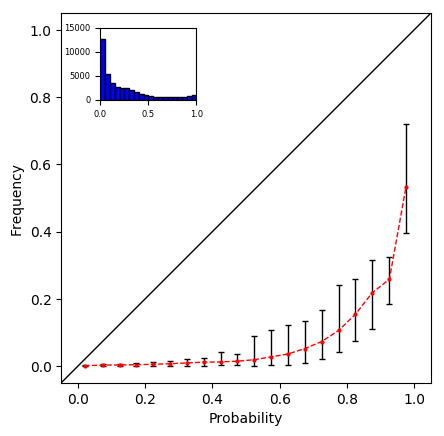}{0.33\textwidth}{(f) LSTM, RD}}
	\caption{Comparison between RF (top) and LSTM (bottom) for $\geq$M-class flare prediction
where the corresponding SSP, ROC and RD are displayed from left to right.}
	\label{fig:Mvisual} 
\end{figure}

Figure \ref{fig:Mvisual} presents SSP, ROC and RD plots 
for RF and LSTM respectively
when the methods are used in
$\geq$M-class flare prediction. 
Refer to the SSP plots.
For RF, the maximum TSS=0.768 is obtained with a probability threshold of 5\%. 
With this threshold, BACC=0.884$\pm$0.003 and HSS=0.262$\pm$0.004.
For LSTM, the maximum TSS=0.792 is obtained with a probability threshold of 60\%.
With this threshold, BACC=0.896$\pm$0.004 and HSS=0.323$\pm$0.030.
The ROC curve of LSTM is slightly better than that of RF. 
LSTM has an AUC of 0.948$\pm$0.003, which is better than RF with an AUC of 0.935$\pm$0.002. 
Refer to the RD plots.
The curve of RF is closer to the diagonal than the curve of LSTM. 
The BS and BSS achieved by RF are 0.021$\pm$0.001 and 0.260$\pm$0.006 respectively. 
The BS and BSS achieved by LSTM are 0.090$\pm$0.009 and -2.241$\pm$0.319 respectively.
These results suggest that RF be a better probabilistic forecasting model than LSTM. 

\begin{figure}
	\gridline{\fig{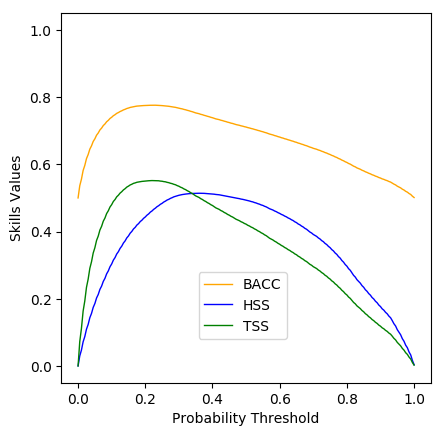}{0.33\textwidth}{(a) RF, SSP}
		\fig{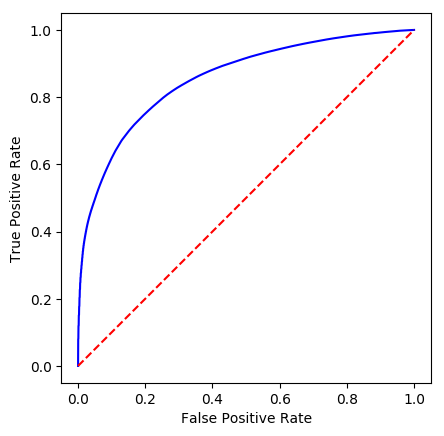}{0.33\textwidth}{(b) RF, ROC}
		\fig{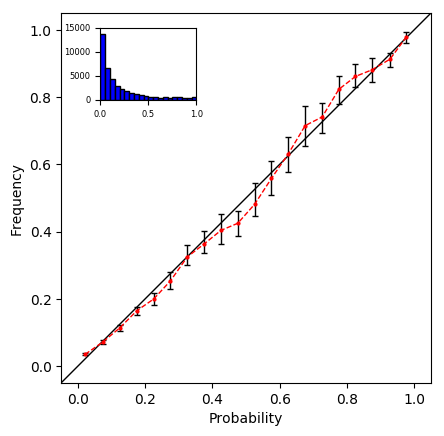}{0.33\textwidth}{(c) RF, RD}}
	\gridline{\fig{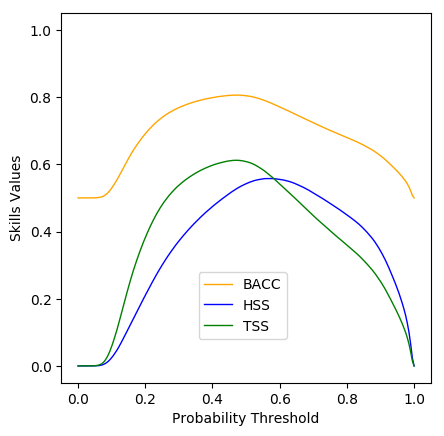}{0.33\textwidth}{(d) LSTM, SSP}
		\fig{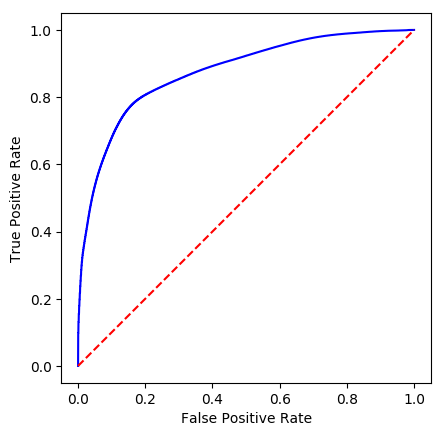}{0.33\textwidth}{(e) LSTM, ROC}
		\fig{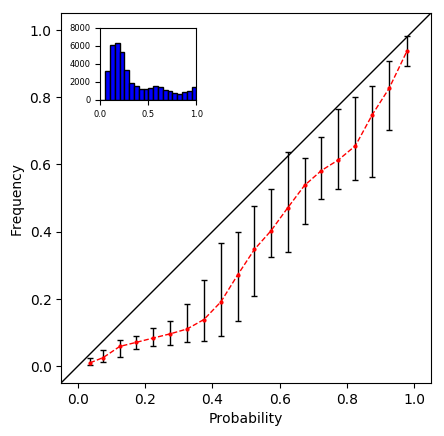}{0.33\textwidth}{(f) LSTM, RD}}
	\caption{Comparison between RF (top) and LSTM (bottom) for $\geq$C-class flare prediction
where the corresponding SSP, ROC and RD are displayed from left to right.}
	\label{fig:Cvisual} 
\end{figure}

Figure \ref{fig:Cvisual} presents SSP, ROC and RD plots 
for RF and LSTM respectively
when the methods are used in $\geq$C-class flare prediction.
Refer to the SSP plots.
For RF, the maximum TSS=0.552 is obtained with a probability threshold of 25\%.
With this threshold, BACC=0.776$\pm$0.001 and HSS=0.469$\pm$0.003.
For LSTM, the maximum TSS=0.612 is obtained with a probability threshold of 50\%. 
With this threshold, BACC=0.806$\pm$0.004 and HSS=0.526$\pm$0.021.
The ROC curve of LSTM is better than that of RF. 
LSTM has an AUC of 0.871$\pm$0.002, which is better than RF with an AUC of  0.851$\pm$0.001. 
Refer to the RD plots. 
The curve of RF almost overlaps the diagonal.
The BS and BSS achieved by RF are 0.103$\pm$0.001 and 0.344$\pm$0.002 respectively. 
The BS and BSS achieved by LSTM are 0.133$\pm$ 0.007 and 0.152$\pm$0.047 respectively.
These results indicate
 that RF is better than LSTM when the methods are used as probabilistic forecasting models.

\section{Discussion and Conclusions}\label{sec:conclusion}

We develop a long short-term memory (LSTM) network to 
predict whether an AR would produce a 
$\Upsilon$-class flare within the next 24 hours.
We consider three $\Upsilon$ classes, namely
$\ge$M5.0 class, $\ge$M class, and $\ge$C class, and
build three LSTM models separately,
each corresponding to a $\Upsilon$ class.
Each LSTM model is used to make predictions of its corresponding $\Upsilon$-class flares.
We build a dataset containing the data in the period from 2010 May to 2018 May,
gathered from the JSOC website. 
Each sample in the dataset has 40 features, 
including 25 magnetic parameters provided by SHARP and related data products 
as well as 15 flare history parameters. 
We divide the dataset into three subsets:
the subset covering 2010--2013 for training,
the subset covering 2014 for validation, and
the subset covering 2015--2018 for testing.
The training subset and testing subset are disjoint, and hence
our proposed method will make predictions on ARs that 
it has never seen before.
With extensive experiments,
we evaluate the performance of all three LSTM models
and compare them with closely related machine learning methods  
using different performance metrics. The main results are summarized as follows.

\begin{quote}
	1. Solar data samples in an AR are considered as time series in this study.
	Although some researchers
	\citep{2017ApJ...835..156N,2018SoPh..293...48J}
	utilize information concerning flaring history for solar flare forecasts,
	none of the previous studies model the data samples as time series and 
	adopt LSTMs to capture dependencies in the temporal domain of the data samples.
	To our knowledge, this is the first attempt of using LSTMs for solar flare prediction.
	
	2. We evaluate the importance of each of the 40 features used in this study. 
	Our experimental results show that, among these 40 features, 
	10 \textit{SDO}/HMI magnetic parameters including
	TOTUSJH, TOTBSQ, TOTPOT, TOTUSJZ, ABSNJZH, SAVNCPP, USFLUX, AREA\_ACR, MEANPOT, TOTFX, 
	and 6 flare history parameters including Cdec, Chis, Chis1d, Edec, Mhis and Xmax1d 
	are more important than the other features for flare prediction.
Our findings on the 
\textit{SDO}/HMI magnetic parameters are
mostly consistent with those reported in
			\citet{2015ApJ...798..135B}. 
It was also observed that the history of C-class flares contributes the most to flare prediction
			among all the flare history parameters.
			Although the rankings of the features are not the same for the three
			LSTM models, some features such as TOTUSJH, TOTUSJZ, TOTPOT, TOTBSQ, USFLUX, SAVNCPP, Cdec, Chis and Chis1d
			exhibit great predictive power for all the three models.
Furthermore, using only 14-22 most important features 
			including both flare history and magnetic parameters
			can achieve better performance than using all the 40 features together. 
		
	3. 
 Our LSTM-based approach
achieves better performance than related
machine learning methods such as 
 multilayer perceptrons (MLP) \citep{haykin2004comprehensive, 2018SoPh..293...28F},
Jordan network (JN) \citep{jordan1997serial},
 support vector machines (SVM) \citep{2007SoPh..241..195Q, 2010RAA....10..785Y, 2015ApJ...798..135B, 2015ApJ...812...51B, 2015SpWea..13..778M, 2018SoPh..293...28F},
and a recently published deep learning-based method, Deep Flare Net \citep[DeFN;][]{2018ApJ...858..113N}.
	In addition, we conduct an ablation study by considering three alternative architectures (ablations) 
	LSTM$_{-a}$, LSTM$_{-c}$ and LSTM$_{-ac}$.
	Our experimental results show that the proposed LSTM architecture achieves better performance 
	than the three ablations,
	demonstrating the effectiveness of adding the attention layer and fully connected layers
	to LSTM units. 

	4. A related machine learning algorithm, namely random forests (RF) \citep{2016ApJ...829...89B, 2017ApJ...843..104L, 2018SoPh..293...28F},
is comparable to our LSTM method.
If one is interested in getting a probabilistic estimate of 
how likely an AR will produce a 
$\ge$M5.0- ($\ge$M-, $\ge$C-, respectively)
class flare within the next 24 hours, 
then RF would be the best choice. 
On the other hand, if one is interested in getting a firm answer regarding whether or not an AR
will produce a 
$\ge$M5.0- ($\ge$M-, $\ge$C-, respectively)
 class flare within the next 24 hours, 
 then our LSTM method is
significantly better than RF
and is recommended.
\end{quote}

Based on our experimental results, we conclude that the proposed LSTM-based framework is a 
valid method for solar flare prediction. 
Considering flare history parameters, besides \textit{SDO}/HMI magnetic parameters,
helps improve prediction performance, a finding consistent with that reported in
\citet{2017ApJ...835..156N} and \citet{2018SoPh..293...48J}.
As solar data from various instruments are gathered at an unprecedented rate,
some researchers attempt to employ additional features including solar images 
for flare prediction \citep{2018SoPh..293...48J}. 
In future work, we plan to incorporate the image data into deep learning methods for predicting
solar flares and other events (e.g., filament eruptions and CMEs).\\

We thank the referee for very helpful and thoughtful comments.
We also thank the team of \textit{SDO}/HMI for producing vector magnetic field data products. 
The X-ray flare catalogs were prepared by and made available through NOAA NCEI.
The related machine learning methods studied in this work
were implemented in Python.
C.L. and H.W. acknowledge the support of NASA under grants  NNX16AF72G, 80NSSC17K0016,
80NSSC18K0673 and 80NSSC18K1705.

\end{document}